
\magnification=\magstep1
\raggedbottom
\parskip 0pt plus0pt minus0pt
\normalbaselineskip 16pt plus0pt minus0pt
\def\section#1{\bigbreak{\noindent{\bf #1}\hfil}\nobreak
   \medskip\nobreak\noindent}
\def\keywords{\medskip\noindent{\bf Key words:\ }}
\def\abstract{\noindent{\bf Abstract.\ }}
\def\acknowledgements{\bigskip\noindent{\it Acknowledgements.\ }}
\def\caption#1{\smallskip\noindent{\bf Fig.\ #1.}}
\def\references{\vfil\eject{\noindent\bf References\hfil}
\par\bigskip\beginrefs}

\def\beginrefs{\begingroup\parindent=0pt\frenchspacing
\parskip=1pt plus 1pt minus 1pt\interlinepenalty=1000\tolerance=400
\hyphenpenalty=10000\everypar={\hangindent=1.6pc}}

\def\endrefs{\endgroup}
\def\ts{\thinspace}


\font\ions=cmcsc10
\def\la{$\lambda$}
\def\ha{H$\alpha$}
\def\hb{H$\beta$}

\def\sline#1#2{#1\ts{\ions#2}}
\def\oiii{[\sline O{iii}]}
\def\oiiia{\oiii\ts\la4959}
\def\oiiib{\oiii\ts\la5007}

\def\heii{\sline{He}{ii}}
\def\heiia{\heii\ts\la4686}
\def\sii{[\sline S{ii}]}
\def\siia{\sii\ts\la6717}

\def\feii{\sline{Fe}{ii}}

\def\feiib{\feii\ts\la5018}

\def\oi{[\sline Oi]}
\def\oia{\oi\ts\la6300}

\def\nii{[\sline N{ii}]}
\def\niia{\nii\ts\la6548}
\def\niib{\nii\ts\la6584}
\def\ergscm{erg s$^{-1}$ cm$^{-2}$}
\def\ergscma{\ergscm\ \AA$^{-1}$}
\def\kms{\hbox {km\ s$^{-1}\,$}}         
\def\deg{\ifmmode {^\circ}                
         \else {$^\circ$}
         \fi
         \hskip -0.3em}
\def\arcsec#1.#2 {\ifmmode {#1^{\scriptscriptstyle\prime\prime}
                            \hskip-0.42em.\hskip0.10em#2}
         \else {$#1^{\scriptscriptstyle\prime\prime}
                     \hskip-0.42em.\hskip0.10em#2$}
         \fi}

\normalbaselines
\def\mkn{Markarian~279}

\null
\vfill
\centerline{\bf Monitoring of active galactic nuclei$^{\star,\star\star}$}
\centerline{\bf V. The Seyfert 1 galaxy Markarian~279}
\bigskip
\centerline{G.M. Stirpe$^1$, D. Alloin$^2$, D.J. Axon$^3$, J. Clavel$^4$,
            A.G. de~Bruyn$^5$, A. del Olmo$^6$,}
\centerline{M. Goad$^7$, P.M. Gondhalekar$^8$, N. Jackson$^9$,
            W. Kollatschny$^{10}$, E. Laurikainen$^{11}$,}
\centerline{A. Lawrence$^{12}$, J. Masegosa$^6$, I.M. McHardy$^{13}$,
            M. Moles$^6$, P.T. O'Brien$^7$,}
\centerline{M.V. Penston$^{14,\dagger}$, J. Perea$^6$, E. P\'erez$^{15}$,
            I. P\'erez-Fournon$^{15}$, A. Robinson$^{16}$,}
\centerline{J.M. Rodr\'\i guez-Espinosa$^{15}$, C.N. Tadhunter$^{17}$,
            R.J. Terlevich$^{14}$, S.W. Unger$^{14}$,}
\centerline{E. van~Groningen$^{18}$, B. Vila-Vilar\'o$^{15}$,
            S.J. Wagner$^{19}$}
\bigskip

\item{$^1$}Osservatorio Astronomico di Bologna, Via Zamboni 33, I-40126
Bologna, Italy

\item{$^2$}Observatoire de Paris, URA173 CNRS, Universit\'e de Paris~7, Place
Jules Janssen, F-92195 Meudon Principal Cedex, France

\item{$^3$}NRAL, Jodrell Bank, Macclesfield, Cheshire SK11 9DL, United
Kingdom

\item{$^4$}ESA Astrophysics Division, ESTEC, Postbus 299, NL-2200 AG
Noordwijk,\hfill\break The Netherlands

\item{$^5$}Radiosterrewacht, Postbus 2, NL-7990 AA Dwingeloo, The
Netherlands

\item{$^6$}Instituto de Astrof\'\i sica de Andaluc\'\i a, CSIC, Apartado 3004,
E-18080 Granada, Spain

\item{$^7$}Department of Physics and Astronomy, UCL, Gower Street, London
WC1E 6BT, United Kingdom

\item{$^8$}Rutherford Appleton Laboratory, Chilton, Didcot, Berkshire OX1
0QX, United Kingdom

\item{$^9$}Sterrewacht Leiden, Postbus 9513, NL-2300 RA Leiden, The
Netherlands

\item{$^{10}$}Universit\"ats-Sternwarte, Geismarlandstra\ss e 11, W-3400
G\"ottingen, Germany

\item{$^{11}$}Turku University Observatory, Tuorla, SF-21500 Piikki\"o, Finland

\item{$^{12}$}Queen Mary College, University of London, Mile End Road, London
E1, United King\-dom

\item{$^{13}$}Physics Department, The University, Southampton SO9 5NH, United
Kingdom

\item{$^{14}$}Royal Greenwich Observatory, Madingley Road, Cambridge CB3 0EZ,
United Kingdom

\item{$^{15}$}Instituto de Astrof\'\i sica de Canarias, E-38200 La Laguna,
Tenerife, Spain

\item{$^{16}$}Institute of Astronomy, Madingley Road, Cambridge CB3 0HA, United
Kingdom

\item{$^{17}$}Department of Physics, Hounsfield Road, Sheffield S3 73H, United
Kingdom

\item{$^{18}$}Astronomiska Observatoriet, Box 515, S-75\ts120 Uppsala,
Sweden

\item{$^{19}$}Landessternwarte, K\"onigstuhl, W-6900 Heidelberg, Germany

\vskip 2truecm

{\parindent=0pt

G.M. Stirpe et al.: Monitoring of active galactic nuclei. V

\medskip

{\it Send offprint requests to:\/} G.M. Stirpe

\medskip

$^\star$ Based on observations made with the William Herschel, Isaac Newton,
and Jacobus Kapteyn Telescopes, operated on the island of La~Palma by the Royal
Greenwich Observatory at the Spanish Observatorio del Roque de los Muchachos of
the Instituto de Astrof\'\i sica de Canarias.

$^{\star\star}$ This article is based on work carried out by the LAG (Lovers of
Active Galaxies) Collaboration. LAG is a consortium of mainly European
astronomers which was established to study active galaxies using the
International Time allocation at the Canary Islands observatories operated
under the auspices of the {\it Comit\'e Cient\'\i fico Internacional\/}.

$^\dagger$ Deceased 23 Dec. 1990

\vfill

\hrule
\medskip

Submitted to: Astronomy \& Astrophysics, Main Journal

Section: Extragalactic astronomy

Thesaurus codes: 11.09.1 Markarian 279; 11.19.1; 02.12.3

Send proofs to: G.M. Stirpe
}

\eject

\abstract We report on the Lovers of Active Galaxies' (LAG) monitoring of the
Seyfert~1 galaxy \mkn\ from January to June 1990. The source, which was in a
very bright state, gradually weakened after the first month of monitoring: the
\ha\ and \hb\ flux decreased by 20\% and 35\% respectively, and the continuum
under \ha\ by 30\%. The luminosity-weighted radius of the broad line region
(BLR), as derived from the cross-correlation function, is of the order of
10~light days. This result is very uncertain because the features in the light
curves are very shallow, but it is unlikely that the radius of the BLR is more
than 1~light month.

The profile variations of \ha\ confirm that the prevailing motions are not
radial. The data of the present campaign and those obtained in previous years,
when the source was in a much weaker state, show that the red asymmetry of the
Balmer lines correlates positively with the broad line flux. This new effect is
briefly discussed.

\keywords Galaxies: individual: Markarian 279 -- Galaxies: Seyfert -- Line:
profiles

\section{1. Introduction}%
The continuum and emission line variability of broad-line active galactic
nuclei (AGN) is a phenomenon which has been known for several decades
(Andrillat \& Souffrin 1968; see also Peterson (1988) for a review of early
results). The typical time scales are of the order of a week or less. Line and
continuum monitoring has been extensively used in recent years as one of the
very few tools available to probe the innermost region ($\ll1$~pc) of AGN,
specifically the spatially unresolved broad line region (BLR). The assumption
is made that the ionizing continuum source is point-like with respect to the
BLR, and that therefore the broad line gas reacts with light travel-time delay
to the variations of the continuum. Simultaneous monitoring of the continuum
and broad emission lines can thus provide information on the structure, size
and kinematics of the BLR, through the technique of reverberation mapping
(Blandford \& McKee 1982). Because of the large amount of telescope time
required to monitor AGN variability, at present only a few objects have been
studied with adequate time resolution and coverage. The most recent results are
reviewed by Peterson (1993). Detailed information on the velocity field of the
BLR is still not available, but data presented by Clavel et al.\ (1990),
Koratkar \& Gaskell (1991) and Stirpe \& de~Bruyn (1991) indicate that
dominant radial motions can be ruled out.

One of the programmes carried out by the Lovers of Active Galaxies (LAG)
collaboration consisted in the frequent high-quality monitoring of a sample of
8~broad-line AGN, spanning a wide range of intrinsic luminosities. The main
purposes of the programme consisted of increasing the sample of well-monitored
AGN both in number and luminosity range, and adding high spectral resolution to
frequent sampling. The observations took place from early January to early
June~1990. In this paper we present the data obtained for one of these sources,
the Seyfert~1 galaxy \mkn, and the immediate conclusions which they imply. For
a review of the previous variability of \mkn, see Stirpe (1991) and references
therein. Jackson et al.\ (1992), Wanders et al.\ (1993), Salamanca et al.\
(1993) and Dietrich et al.\ (1993a) discuss results from other targets of the
LAG campaign.

\section{2. Observations and reduction}%
The observations of \mkn\ during the LAG campaign were conducted with the
double spectrograph ISIS at the 4.2m William Herschel Telescope (WHT), and with
the Intermediate Dispersion Spectrograph (IDS) at the 2.5m Isaac Newton
Telescope (INT). CCDs were attached to both spectrographs. Tables 1 and~2 list
the observations for \ha\ and \hb\ respectively. For each observation, a letter
(A--D) indicates the instrumental configuration used, following the definitions
in Table~3. A wavelength interval of $\sim800$~\AA\ covering \ha\ was observed
at all epochs except the last. When enough time was available a similar
interval covering \hb\ was also observed. A slit width of \arcsec1.5 was used
throughout the campaign. The slit was oriented in the NS direction.

Standard procedures were followed to reduce the spectra, using the Starlink
software package Figaro (Fuller 1989). The mean bias level was measured on the
overscan section of each CCD frame, and a corresponding constant was subtracted
from the entire frame. The pixel-to-pixel sensitivity variations were then
corrected for by dividing each frame by a tungsten lamp flat-field (normally
taken at the beginning or end of the same night) which had previously been
normalized in the wavelength direction. Synthetic sky frames were obtained by
fitting a 3rd order polynomial to each row (i.e.\ in the spatial direction) of
a 2-D spectrum, discarding from the fit all pixels corresponding to bad
columns, cosmic ray events, and target spectrum. Each fitted sky frame was
subtracted from its corresponding spectrum frame.

Cu/Ar and Cu/Ne lamps were used for the wavelength calibration of the \hb\ and
\ha\ regions respectively. Depending on the resolution and number of useful
lines in the arc frame (always more than 10, and occasionally exceeding 20),
2nd to 5th order polynomial fits were used, which yielded residuals lower than
0.4~\AA\ at the lowest resolution and lower than 0.1~\AA\ at the highest
resolution.

All spectra were corrected for atmospheric extinction using the coefficients
provided by the Royal Greenwich Observatory for La~Palma. The flux calibration
was achieved with the stars HD~84937 and BD~+26\deg2606, using the fluxes
tabulated by Oke \& Gunn (1983). A correction for the B-band of O$_2$ was
obtained by interpolating the continuum of the same stars, and used to divide
the band out of the \ha\ spectra of \mkn. The spectra of \mkn\ were then
extracted by summing a number of spectrum columns corresponding to the same
spatial interval (\arcsec6.5) on all frames. Finally, the wavelength scale was
corrected for the redshift of \mkn\ ($z=0.0303$): unless otherwise stated, all
wavelengths mentioned hereafter are in the rest system of \mkn.

Tables 1 and 2 also list the spatial full width at half maximum (FWHM) of the
unresolved portion of each spectrum. This quantity was obtained from each frame
as follows. A 2-D section of the frame containing only continuum (underlying
galaxy + non-stellar component) and one containing also the broad emission line
were collapsed in the wavelength direction. The spatial profiles thus obtained
present extended wings formed by the resolved underlying galaxy. The continuum
profile was scaled so that its extended wings had the same flux as those of the
continuum + broad line profile, and the former was then subtracted from the
latter. The positive residual obtained was assumed to be emitted entirely by
the unresolved BLR. The cores of the residual profiles could be well fitted by
Gaussians: the FWHM of the fits are listed in Tables 1 and~2. These values
provide an indication of the seeing for each observation.

\section{3. Derivation of light curves}%
The spectra were corrected for differences in flux scale (caused by differences
in seeing and transparency) by applying the internal calibration discussed in
Stirpe \& de~Bruyn (1991), based on the assumption that the narrow lines do not
vary. The lines used as calibrators were the strongest of each wavelength
region, i.e.\ \oiiia\ and \la5007 for \hb, and the narrow \ha\ and \niib\ for
\ha. The spectra obtained on 31~January (JD2447923) were used as reference for
the scaling. The fluxes of the \oiiib, narrow \ha, and \niib\ lines in these
spectra are $(14.3\pm0.8)\times10^{-14}$ \ergscm, $(6.2\pm0.9)\times10^{-14}$
\ergscm, and $(6.2\pm0.9)\times10^{-14}$ \ergscm\ respectively. The
uncertainties were estimated from the rms of the same measurements made on all
spectra obtained on photometric nights with good seeing ($\le\arcsec1.5 $).

The internal calibration method consists in scaling a spectrum with respect to
the reference spectrum, and adjusting the scaling factor until the narrow line
residuals in their difference are minimized. The residuals were minimized
automatically with the algorithm described by van~Groningen \& Wanders (1992).
This procedure is more objective and accurate than visual inspection, and led
to uncertainties in the relative calibration of typically $\le3$\% for \ha\ and
$\sim2$\% in \hb. The uncertainties were estimated by perturbing the internal
calibration of each spectrum until clear narrow line residuals appeared in the
difference between it and the reference spectrum. The narrow line region (NLR)
of \mkn\ is compact and unresolved, and therefore the internal calibration is
not affected by narrow line slit losses as, for instance, NGC~3516 (Wanders et
al.\ 1993). Figures 1 and~2 show the scaled \ha\ and \hb\ spectra.

A power-law continuum was fitted to all \ha\ spectra to intervals of 40~\AA\
centred at 6140~\AA\ and 6820~\AA. For the spectra obtained on 16~February
(JD2447939) and 17~March (JD2447968), whose wavelength ranges are slightly
shifted bluewards, the second interval was centred at 6800~\AA\ and 6780~\AA\
respectively, and for the latter it was narrowed to 10~\AA. After the continuum
subtraction, the broad component of \ha\ was isolated by subtracting the narrow
\ha\ component, and the forbidden lines \oia\ and \la6364, \siia\ and \la6731,
and \niia\ and \la6584. The \oiiib\ profile was used as a template for the
narrow \ha, and high signal-to-noise templates for the other forbidden lines
were obtained from the average of all \ha\ spectra. To isolate \niib\ in the
average spectrum, a smooth broad \hb\ template was subtracted from \ha\ after
the subtraction of the narrow Balmer components, in order to eliminate the
first order structure underlying the \nii\ lines. The \nii\ template thus
obtained was used also for \niia, with the appropriate intensity scaling.
Figure~3 shows the \ha\ spectrum of 31~January before and after deblending. The
broad line components resulting from this cleaning process were integrated
between 6400~\AA\ and 6750~\AA\ to obtain the light curve of \ha.

The broad \hb\ profiles were obtained as follows. A broad \ha\ template was
used to subtract the multiplet~42 \feii\ lines (\la4924 and \la5018) and
\heiia\ from the spectra. The \oiii\ template was used to deblend \hb\ from its
own narrow component and from \oiiia\ and \la5007. Finally, a power-law
continuum was fitted to and subtracted from the results. The \hb\ spectrum
obtained on 31~January is shown in Fig.~4, before and after the deblending. The
deblending of \hb\ is inevitably a subjective procedure, because of the
contamination of its wings by the \feii\ and \heii\ lines. The highest
uncertainty affects the placing of the continuum. In order to derive a
realistic error bar for the broad \hb\ fluxes, the deblending, continuum
subtraction, and flux measurements were repeated with ``extreme" continua
(i.e.\ continua considered to represent upper or lower limits to the true
ones): the relative uncertainty of the fluxes thus obtained is $\sim5$\%. This
does not include systematic uncertainties caused by the choice of the
continuum-fitting method. The \hb\ fluxes were obtained by integrating the
deblended spectra between 4760~\AA\ and 4960~\AA.

The data-set formed by the \ha\ spectra is more complete than that formed by
the \hb\ spectra. Therefore the non-stellar continuum light curve had to be
derived from the former. The continuum at \ha\ was represented by the values
of the power-law fits at the rest wavelength of \ha. These were corrected for
seeing effects as described in the Appendix. The use of such a red continuum
has a drawback: it is likely in fact that the amplitude of the variations is
much smaller in the red continuum with respect to the ionizing continuum, or
even with respect to the optical blue continuum, given that Seyfert variability
tends to be stronger at higher frequencies (e.g.\ Peterson et al.\ 1991).
Furthermore, the stellar continuum under \ha\ causes additional dilution of the
variations. On the other hand, there is no evidence that the optical continuum
is delayed with respect to the UV by more than 2 days in NGC~5548, a Seyfert~1
of intrinsic luminosity comparable to that of \mkn\ (Peterson et al.\ 1991),
and from one of the best studied cases up to now (Fairall~9, Clavel et al.\
1989) it appears that the delay effects observed in the IR continuum with
respect to the optical and UV continua do not extend to the red part of the
optical spectrum. We therefore assume that the light curve of the red continuum
is simply a lower amplitude representation of those of the UV and blue
continuum.

The measured and corrected continuum fluxes (F$_{\lambda6563}$ and
F$_{\lambda6563}^{corr}$ respectively) and the integrated broad \ha\ and
\hb\ fluxes are listed in Tables 4 and~5, and the light curves are shown
in Fig.~5. The intervals between consecutive observations of the \ha\ region
range from 1 to 18 days, with an average value of 6.1~days.

\section{4. Continuum and line flux variations}%
When the monitoring started in January 1990, \mkn\ was in a much brighter state
than in previous years: the broad \ha\ component in the LAG spectra is about
twice as strong when compared with the data obtained in 1987 by Stirpe \&
de~Bruyn (1991) and in 1988 by Stirpe \& de~Bruyn (in preparation) and Maoz et
al.\ (1990). Its intensity is comparable to that observed before the low state
(e.g.\ Stirpe, 1990).

After a short quiescent period at the beginning of the campaign, the broad \ha\
flux increased by 10\%, and after about 20~days decreased again by the same
amount, and continued to decrease slowly until the end of the monitoring
season. A shallow secondary maximum is visible around JD2447995. There is no
isolated maximum in the continuum light curve corresponding to that visible in
the \ha\ curve between JD2447921 and JD2447940: lack of observations and
scaling uncertainties may have hidden a driving feature in the continuum curve
corresponding to the \ha\ maximum. The peak-to-peak variation of \ha\ during
the entire campaign did not exceed 20\%. The continuum underwent a higher
amplitude decrease ($\sim 30\%$), and only very shallow deviations from the
generally negative slope of the light curve are visible. Notice that the
apparently correlated short time scale, low amplitude variations in the two
light curves are most probably due to scaling uncertainties, both curves being
derived from the same spectra. The \hb\ light curve confirms the decreasing
trend, with an amplitude of 35\% (higher than that of the \ha\ variations, as
is usually observed in varying Seyfert~1 nuclei). This light curve is too
sparsely sampled to add significant information to the \ha\ data. The \heiia\
and \feii\ lines also decreased during the campaign.

CCD images of \mkn\ were obtained for the duration of the LAG campaign at the
1m Jacobus Kapteyn Telescope (JKT). The nuclear fluxes obtained from the images
are not accurate enough to confirm or add information to the spectroscopic
results, but are consistent with them within the uncertainties. The fluxes
decreased by $\le50$\% and $\le40$\% in the {\it B\/} and {\it V\/} bands
respectively, between mid-January (JD2447910) and late May (JD2448040). The
variability in the {\it R\/} and {\it I\/} bands was $\le20$\%.

Figure 6 shows the cross correlation function (CCF) of the \ha\ vs. continuum
light curves, and the autocorrelation function (ACF) of the continuum light
curve, calculated with the Gaskell \& Peterson (1987) method, but without
artificially extending the light curves with a constant value before and after
the campaign (in other words, for each lag only the overlapping branches of the
light curves were cross correlated). The fluxes obtained on JD2447975, which
have the highest scaling uncertainties because of the low signal-to-noise ratio
in the corresponding spectrum, were eliminated from both light curves before
computing the two functions. The sampling window ACF (see Gaskell \& Peterson
1987), also shown in Fig.~6, has $\rm FWHM \le 6$~days, implying that the
variations on longer time scales are resolved.

The ACF is not only unusually broad (it crosses the zero line at $\pm$52~days,
and declines very slowly from its maximum), but is also broader than the CCF.
This would imply that \ha\ varied on shorter time scales than the continuum,
which is inconsistent with a simple reverberation scenario, or that the
transfer function of \ha\ becomes negative at short lags (Sparke 1993, Goad et
al.\ 1993). Yet another possibility is that the continuum variations are
undersampled: this is however unlikely, given that the sampling window ACF is
much narrower than the continuum ACF. The discrete cross correlation function
(DCF), calculated as described by Edelson \& Krolik (1988) with the
modifications described in Reichert et al.\ (1993), is also plotted in Fig.~6,
and is consistent with the CCF. Pairs of points obtained from the same spectra
were not used in the calculation of the DCF, to eliminate the effect of
correlated calibration errors at zero lag.

The peak value of the CCF is at a lag of 6~days, and the DCF reaches its
maximum in the bin centred on the same lag. P\'erez et al.\ (1992a) have
cautioned against using the peak value of the CCF as an indication of
the size of the BLR, particularly when the time baseline does not cover several
years. In this case the width of the ACF, the presence of correlated errors at
zero lag, and the short duration of the campaign all weigh on the estimates
provided by the CCF. Furthermore, the result is strongly influenced by the
first three epochs, in which the continuum appears to be at the highest level
of the campaign while the line still has to reach its maximum. If these epochs
are not included when calculating the CCF, the resulting lag is zero. It is
therefore likely that the peak value of the CCF is only a poor indication of
the inner radius of the BLR, even beyond the formal error of 4.5 days (Gaskell
\& Peterson 1987). Some indication on the uncertainty of the peak value can be
obtained by using the method described by Maoz \& Netzer (1989) which consists
in sampling the interpolated light curves many times with the same amount of
observations in different patterns, adding random noise comparable to the
uncertainties of the measurements, and obtaining the resulting cross
correlation peak distribution (CCPD). In our case 90\% of the values in the
CCPD are between $-1$ and $+22$~days, and 50\% between 3 and 15~days. This
confirms the high uncertainty; however, there is no evidence that the lag is
longer than about 1~month.

The centroid of the CCF is at 10.2~days: this can be interpreted as an
estimate of the luminosity-weighted radius of the BLR (Robinson \& P\'erez
1990, Koratkar \& Gaskell 1991), but again caution must be used and this number
should be considered only as a rough estimate (P\'erez et al.\ 1992a).

\section{5. Line profile variations}%
The broad line profiles of \mkn\ underwent some variation during the LAG
campaign: during the first half of the campaign the red side of \ha\ was
clearly more convex than in the second half (Fig.~1). Figure~7, which shows the
difference spectra between high and low state spectra of both \ha\ and \hb,
together with the high state broad profiles, demonstrates how the red side of
the lines varied more strongly than the blue side. The peaks of the lines in
the difference spectra are shifted to $\sim1000$~\kms. This behaviour does not
generalize to the entire observational history of \mkn, as Peterson et al.\
(1982) reported stronger variations of the blue side of \hb\ between March and
May 1981.

The profile changes observed during the LAG campaign are closely related to the
luminosity of the line: this is evident from Fig.~8, which shows the fraction
of broad \ha\ flux on the red side of the rest wavelength, plotted against the
total line flux. The plot also includes measurements from spectra taken at the
INT in previous years, with the same instrumental configurations used for the
LAG data (most of the spectra used are published in Stirpe 1990, 1991 and
Stirpe \& de~Bruyn 1991). All the spectra were internally calibrated to a
common scale, and the broad components were isolated in a consistent manner.
There is obviously a correlation between the two quantities plotted in Fig.~8,
in the sense that the line is more red-asymmetric when it is stronger. The
ordinate in Fig.~8 is only a rough indicator of the shape of the line; however,
a comparison between pairs of spectra corresponding to points which are close
in the diagram but separated by several weeks or months in time (e.g.\
JD2447894 and JD2447968, JD2447897 and JD2447995, JD2447911 and JD2447944)
shows that the shapes are identical within the noise.

Because the \ha\ spectra were internally calibrated by minimizing the residuals
of narrow lines which are superimposed on the broad \ha\ component, there is a
possibility that the shape of the broad line at a given epoch influences the
internal calibration, causing a spurious correlation in the asymmetry vs.\ line
flux diagram. This, however, is ruled out by the wide range of broad \ha\
fluxes: between July~1987 and February~1990 the line doubled in intensity, and
the uncertainty introduced by the internal calibration is much lower than the
total range covered by the variability. Further confirmation of the effect
comes from Fig.~9, which shows the same diagram as Fig.~8 derived for \hb.
Within the LAG data-set no correlation is visible, which is not surprising
given the large uncertainties involved when isolating \hb\ from the underlying
continuum and from the contaminating \feii\ and \heii\ lines. However, the
points obtained from previous spectra, some of which were taken during the
source's low state, confirm the correlation found for \ha. Notice that in this
case the internal calibration is based on a strong line (\oiiib), which hardly
overlaps with the \hb\ profile. Although the \feiib\ line underlies \oiiib, the
latter's residuals in the difference spectra are narrow enough to be minimized
without being affected by the presence of a relatively weak broad line. We
therefore conclude that the strength or shape of the broad lines does not
influence the internal calibration of the \hb\ spectra.

A closer inspection of the \ha\ profiles reveals that most of the profile
variation is in fact confined to the red side of the line. As an example,
Fig.~10 shows the broad \ha\ profile in its highest state (observed on 8
February 1990) compared with the average profile of the 1987 campaign by
Stirpe \& de~Bruyn (1991). The latter is scaled by a factor 2.13, in order to
normalize the flux of its blue side to that of the stronger spectrum. The plot
and the accompanying ratio between the two lines show that the blue sides
of \ha\ overlap very well, while the red side becomes much more convex at high
flux. Most of the profile change has occurred between 0~\kms\ and +4000~\kms.
All the \ha\ spectra used for Fig.~8 show a similar effect.

There is no obvious explanation for the correlation between line asymmetry and
line luminosity. A purely radial velocity field can reproduce it if (a) the
continuum light curve has a monotonic trend on time scales longer than the
light crossing time, or (b) the time scales of the continuum fluctuations are
much shorter than the light crossing time. However, the light curves did not
have a monotonic trend during the 5~years separating the first and last points
in Fig.~8, and if (b) were true the BLR would have to be very large and strong
delay effects would be observed between the red and blue sides of the line.
There is however no observable delay between the variations of the two sides of
\ha\ in \mkn, just a difference in their amplitude (Fig.~11). In particular,
the development of the red feature evidenced by Fig.~10 does not display any
delay effect with respect to the main body of \ha. This implies that, while
regions of positive and negative velocity must be distributed fairly evenly
around each iso-delay surface within the BLR (thus excluding pure radial inflow
or outflow), the receding and approaching fractions of the gas react
differently to the continuum variations or, in other words, have different
distributions of physical conditions.

None of the other varying sources monitored by LAG displayed a correlation
between the profile and intensity of \ha, and the effect has been observed
only in two other AGN. One is NGC 5548, whose intrinsic luminosity is
about half that of \mkn: however, as noticed by Rosenblatt \& Malkan (1990),
and confirmed by the analysis presented by Stirpe (1993), the effect in this
source is {\it opposite}, i.e.\ the blue side of the Balmer lines becomes
relatively stronger at higher line fluxes. This may be caused either by very
different structures of the BLRs of \mkn\ and NGC~5548, or by different
orientations of anisotropic BLRs. The only other AGN whose lines are known to
display a profile-intensity correlation is the QSO OQ~208 (Marziani et al.\
1993), which is more luminous than \mkn\ by about a factor~3; in this case also
the peak velocity of the strong red shoulder in the Balmer lines, as well as
its relative strength, correlates with the line luminosity. The correlation was
observed in OQ~208 on a time scale of 6~years.

\section{6. Conclusions}%
We have presented observations of \mkn\ obtained during the LAG monitoring
campaign of January--June 1990. The main results can be summarized as follows.

The light curves of the broad \ha\ line and of its underlying continuum display
variations of no more that 20\% and 30\% respectively (but the continuum
variation is diluted by the host galaxy's spectrum). It appears that, while
\mkn\ has always displayed some variability when monitored in recent years, the
strongest continuum variations occur on long time scales. The limited time
scale of our campaign and the presence of only shallow features in the
continuum and \ha\ light curves do not allow us to constrain the size of the
BLR strongly. We obtain a formal lag of $6\pm4.5$~days from the
cross-correlation analysis. Maoz et al.\ (1990) obtained a lag of 12~days from
a
monitoring campaign conducted in 1988: both results are affected by strong
uncertainties, and should not be considered inconsistent with each other. The
CCPD indicates an upper limit to the \ha\ vs. continuum lag of about 1~month.
The BLR of \mkn, therefore, does not appear to be significantly larger than
that of NGC~5548 (whose \ha\ vs. continuum lag is about 20~days; Dietrich et
al.\ 1993b). The centroid of the CCF yields a rough estimate of 10~light days
for the luminosity-weighted radius of the BLR.

The line variations yield insights on the physical structure of the
BLR. The relation between line profile and line luminosity discussed in Sect.~5
implies that the 2-D transfer function (Welsh \& Horne 1991, P\'erez et al.\
1992b) of the Balmer lines is illumination-dependent in part of the projected
velocity range. On the other hand, the fact that the line profile is correlated
with the luminosity on time scales of several years implies that the transfer
function is stable. In other words, the physical structure of the BLR in \mkn\
does not change on time scales comparable to the dynamical time scale. The line
profile-intensity relation, however, does not appear to generalize to the other
varying objects of the LAG campaign. It may be no coincidence that the relation
is observed in the only object of the LAG sample with medium intrinsic
luminosity: the more luminous objects did not vary enough to reveal the effect,
if it is present, and the BLR of the less luminous objects may be too small for
the effect to be detected. A search for the relation in existing data sets of
AGN of different luminosities would help to determine whether the effect is
common, or whether it is a characteristic of only a few individual AGN. In
addition, further monitoring of \mkn\ and NGC~5548 (not necessarily on short
time scales) will allow us to determine whether the effect persists on time
scales much longer than the dynamical time scale of the BLR, and therefore
whether or not structural changes occur in this region.

\acknowledgements
We are grateful to the staff of the Roque de Los Muchachos Observatory at
La~Palma, for the support provided during the monitoring programme. The CCI is
thanked for the allocation of the international observing time. We acknowledge
the generous hospitality of the Instituto de Astrof\'\i sica de Canarias, whose
computing facilities were used heavily for part of the data reduction, and the
financial support of several national funding agencies, including CNRS, DFG
through grant Ko 857/13-1, ESA, NFR, and SERC. GMS is grateful to Brad Peterson
for providing the most recent version of the Gaskell-Peterson cross-correlation
code.
\vfill\eject

\font\csc=cmcsc10
\def\fwhm{\hbox{\csc fwhm}}

\section{Appendix. Seeing-dependent correction of the continuum light curve}%
The large scatter in the spatial FWHM of the spectra (Table~1) causes the
unresolved non-stellar continuum and the extended underlying galaxy to
contribute in different proportions to each spectrum. In particular, spectra
obtained in bad seeing conditions have a relatively stronger continuum
contribution from the host galaxy after the internal calibration. The amount of
stellar light which contaminates the \ha\ spectra cannot be determined on the
basis of the equivalent width of absorption features (Wanders et al.\ 1993),
because none are visible in the observed spectral range. Therefore, an
approximate correction for seeing effects was derived by using one of the
sharpest {\it R\/} images obtained at the JKT during the LAG campaign (see
Sect.~4). The FWHM of the stellar images on this frame , which was taken on
29~April 1990, is \arcsec1.2. The wavelength range of the {\it R\/} filter
almost exactly overlaps that of the \ha\ spectra, so that we can safely assume
that in both cases the same spectral region was observed. We also assume that
the efficiency of the detector and the transmission of the {\it R\/} filter are
constant across the entire band. This is not strictly true, particularly as far
as the filter transmission is concerned (Unger et al.\ 1988), but the
corrections which we derive are small enough to justify the approximation.

A set of synthetic images with the same spatial FWHM (i.e.\ seeing) as the
spectra (see Table~1) was obtained by convolving the {\it R\/} frame with a 2-D
Gaussian of appropriate width. The flux of \mkn\ was measured within a box of
$\arcsec1.5 \times\arcsec6.5 $ in the EW and NS directions respectively,
corresponding to the widths of the spectrograph slit and extraction window. The
same was done for a comparison star on the same frame. If $F_{M279}(\fwhm)$ and
$F_*(\fwhm)$ represent the fluxes of \mkn\ and of the comparison star
respectively, measured in the chosen window on a frame of given FWHM, and
FWHM$_0$ is the FWHM of the point-like sources in the non-convolved image, then
the quantity
$$C(\fwhm)={F_{M279}(\fwhm) / F_*(\fwhm) \over F_{M279}(\fwhm_0) /
F_*(\fwhm_0)}
- 1$$
yields the fraction of the relative flux of \mkn\ measured on a convolved
image which is in excess of that obtained from the non-convolved image.
Because we assume that the sources of non-thermal continuum and emission lines
have the same point spread function as the comparison star, we attribute this
excess entirely to an increase of the relative contribution of the underlying
galaxy. In other words, the function $C(\fwhm)$ indicates how the stellar
continuum of an internally calibrated spectrum increases as the seeing
increases from \arcsec1.2. The relative flux of \mkn\ within the box
increases by $\sim 10$\% at the worst seeing of the spectroscopic campaign
(\arcsec3.6), with respect to the relative flux measured in the same box on the
non-convolved image.

The $C(\fwhm)$ factors can be transformed in actual fluxes by using the
internally calibrated \ha\ spectrum taken on 2~May~1990, the epoch closest to
that of the image. If $\bar F$ is the average flux of the spectrum over its
entire wavelength range, and if FWHM$_{sp}$ is its spatial width (in this case
\arcsec1.3), the calibrated corrections are
$$C_n(\fwhm) = {\bar F \over 1+C(\fwhm_{sp})}\, C(\fwhm). $$
This requires the assumption that the underlying stellar spectrum is constant
over the entire wavelength range of the spectra. Because this range is narrow
($\sim$800~\AA), the assumption is acceptable.

All continuum fluxes derived from the spectra were corrected by subtracting the
$C_n$ corresponding to their FWHM. The FWHM of some spectra are lower
than that of the image used for the simulations: the corresponding continuum
measurements, therefore, required a positive instead of a negative correction.
The quantity to add (the stellar component `deficit') was derived by
extrapolating the $C(\fwhm)$ curve to the lower FWHM with a low-order
polynomial fit. Although the uncertainty on this derived correction will be
rather high, the correction itself is small enough with respect to the
continuum fluxes to make its uncertainty negligible with respect to that of
the internal calibration.

To check the accuracy of the corrections, an {\it R\/} image obtained at
another
epoch (10~March~1990), also with a FWHM of \arcsec1.2, was used in combination
with the \ha\ spectrum of 17~March~1990 to derive another correction curve with
the same method described above. The two $C_n$ series agree within 10\%, and
the corrected continuum fluxes within 1\%. A further endorsement for the
applied corrections comes from the fact that, while the uncorrected continuum
fluxes correlate weakly with the seeing, no such correlation is present once
the fluxes have been corrected. Finally, the continuum light curve appears
considerably smoother after the corrections are applied.

Note that this procedure does {\it not\/} eliminate the stellar contribution
from the continuum light curve. It simply adjusts all the points in the curve
so that they are in principle contaminated by the {\it same amount\/} of
stellar light: we therefore assume that the variations in the continuum light
curve are caused only by the intrinsic variability of the non-stellar
source, albeit diluted by a constant stellar component. The method as described
can only be applied if the NLR is unresolved, because an extended NLR would
also contribute to $C(\fwhm)$.

\references

Andrillat Y., Souffrin S., 1968, Astrophys.\ Lett. 1, 111\par
Blandford R.D., McKee C.F., 1982, ApJ 255, 419\par
Clavel J., Wamsteker W., Glass I., 1989, ApJ 337, 236\par
Clavel J., Boksenberg A., Bromage G.E., et al., 1990, MNRAS 246, 668\par
Dietrich M., Kollatschny W., Alloin D., et al., 1993a, A\&A (submitted)\par
Dietrich M., Kollatschny W., Peterson B.M., et al., 1993b, ApJ 408, 416\par
Edelson R.A., Krolik J.H., 1988, ApJ 333, 646\par
Fuller N.M.J., 1989, Starlink User Note 86, Rutherford and Appleton
Laboratory\par
Gaskell C.M., Peterson B.M., 1987, ApJS 65, 1\par
Goad M.R., O'Brien P.T., Gondhalekar P.M., 1993, MNRAS 263, 149\par
Jackson N., O'Brien P.T, Goad M., et al., 1992, A\&A 262, 17\par
Koratkar A.P., Gaskell C.M., 1991, ApJS 75, 719\par
Maoz D., Netzer H., 1989, MNRAS 236, 21\par
Maoz D., Netzer H., Leibowitz E., et al., 1990, ApJ 351, 75\par
Marziani P., Sulentic J.W., Calvani M., et al., 1993, ApJ 410, 56\par
Oke J.B., Gunn J.E., 1983, ApJ 266, 713\par
P\'erez E., Robinson A., de la Fuente L., 1992a, MNRAS 255, 502\par
P\'erez E., Robinson A., de la Fuente L., 1992b, MNRAS 256, 103\par
Peterson B.M., 1988, PASP 100, 18\par
Peterson B.M., 1993, PASP 105, 247\par
Peterson B.M., Foltz C.B., Byard P.L., Wagner R.M., 1982, ApJS 49, 469\par
Peterson B.M., Balonek T.J., Barker E.S., et al., 1991, ApJ 368, 119\par
Reichert G.A., Rodr\'\i guez-Pascual P.M., Alloin D., et al., 1993, ApJ
(in press)\par
Robinson A., P\'erez E., 1990, MNRAS 244, 138\par
Rosenblatt E.I., Malkan M.A., 1990, ApJ 350, 132\par
Salamanca I., Alloin D., Baribaud T., et al., 1993, A\&A (in press)\par
Sparke L.S., 1993, ApJ 404, 570\par
Stirpe G.M., 1990, A\&AS 85, 1049\par
Stirpe G.M., 1991, Variability of Markarian 279. In: Duschl W.J., Wagner S.J.,
Camenzind M. (eds.) Variability of Active Galaxies. Springer-Verlag,
Heidelberg, p.~71\par
Stirpe G.M., 1993, A relation between the Profiles and Intensities of Broad
Emission Lines. In: Robinson A., Terlevich R.J. (eds.) The Nature of Compact
Objects in AGN. Cambridge University Press, Cambridge (in press)\par
Stirpe G.M., de Bruyn A.G., 1991, A\&A 245, 355\par
Unger S.W., Brinks E., Laing R.A., Tritton K.P., Gray P.M., 1988, Isaac Newton
Group, La~Palma, Observers' Guide\par
van Groningen E., Wanders I., 1992, PASP 104, 700\par
Wanders I., van Groningen E., Alloin D., et al., 1993, A\&A 269, 39\par
Welsh W.F., Horne K., 1991, ApJ 379, 586
\endrefs
\vfill\eject

\section{Figure captions}
\caption1 The \ha\ spectra of \mkn\ obtained during the 1990 LAG campaign,
after the internal calibration. Each spectrum has an offset of 10 units with
respect to the one below it

\caption2 The \hb\ spectra of \mkn\ obtained by LAG in 1990, each with an
offset of 10 units with respect to the one below it

\caption3 The \ha\ spectrum obtained on 31 January 1990 (JD2447923), before
and after the subtraction of a power-law continuum and of templates for the
narrow \ha\ component, for \oia\ and \la6364, for \niia\ and \la6584, and for
\siia\ and \la6731. The wavelength scale refers to the rest system of \mkn

\caption4 As Fig.~3, for the \hb\ spectrum obtained on the same date. The
deblended line results from the subtraction of a power-law continuum, and of
templates for the narrow \hb\ component, for \heiia, for \oiiia\ and \la5007,
and for the m42 lines of \feii\ (\la4924, \la5018 and \la5169). Notice the
broad wing on the red side of \hb, which cannot be explained as a residual of
\feii\ lines

\caption5 The light curves of (from top to bottom) the continuum under \ha,
corrected for seeing effects as described in the Appendix, the integrated broad
component of \ha, and the integrated broad component of \hb

\caption6 The cross-correlation and discrete cross-correlation functions (CCF
and DCF respectively) of broad \ha\ versus continuum, the auto-correlation
function (ACF) of the continuum, and the sampling window ACF (SWACF) of the
\ha\ spectra. The DCF was calculated in 6-day bins. The light curves shown in
Fig.~5 were used, excluding from both the point with the highest uncertainty
(JD2447975)

\caption7 The difference between high and low state spectra of \ha\ and \hb:
the dates of the individual spectra are given in the figure. No deblending or
continuum subtraction was performed on the spectra used to derive these
differences. The dashed lines show scaled and smoothed versions of the high
state deblended profiles (see Figs.~3 and~4), with added straight line continua
which roughly match the continua in the difference spectra. Notice how the
difference spectra peak at $\sim1000$~\kms, unlike the individual profiles
which peak around 0~\kms. The difference spectrum of \hb\ also evidences the
variation of \heiia\ and \feiib, which show up as two bumps at
$v\sim-11000$~\kms\ and $\sim+10000$~\kms\ respectively; neither is present in
the dashed line profile, because of the deblending described in Sect.~3. The
very broad wing on the red side of \hb\ (see Fig.~4) does not appear in the
difference spectrum

\caption8 The diagram shows, for each \ha\ spectrum of \mkn, the ratio between
the flux of the red side of the line (F(H$\alpha_r$) is the broad \ha\ flux
integrated at $\lambda>6563$~\AA) and the total broad \ha\ flux F(\ha), plotted
against F(\ha) itself: the ordinate is thus a rough indicator of the line
asymmetry. The spectra used for the measurements were obtained in different
observing seasons, as indicated by the symbols

\caption9 As Fig.~7, for \hb; F(H$\beta_r$) is the broad \hb\ flux integrated
at $\lambda>4861$~\AA

\caption{10} The top panel shows the profile of the broad \ha\ component on 8
February 1990 (JD2447931), when the line was in its brightest state during the
LAG campaign, and the average of the profiles from the spectra obtained by
Stirpe \& de Bruyn (1991) in 1987, when \mkn\ was twice as faint. The 1987
profile has been scaled upwards so that the blue sides match as closely as
possible. The excess of the 1990 profile on the red side is $\sim6\%$ of the
broad line flux. Bottom panel: the ratio between the profiles shown in the
upper panel, which evidences how the shape of the red side has changed while
the shape of the blue side has remained almost constant. Notice that the
deviation from 1 at velocities higher than +5000~\kms\ can be attributed to
contamination by residuals of the atmospheric B-band of O$_2$ and of the \siia\
line

\caption{11} The light curves of the red (continuous line) and blue (dashed
line) sides of \ha, each normalized to its mean value. The fluxes of JD2447975,
which have the highest uncertainties, have not been included. The two light
curves show the same features without any relative delay, but the red side
varies with a higher relative amplitude

\vfill\eject

\null\vfill
\centerline{{\bf Table 1.} Journal of observations for \mkn: \ha\ spectra}
\medskip
$$\vbox{
\newdimen\digitwidth
\setbox0=\hbox{\rm0}
\digitwidth=\wd0
\catcode`?=\active
\def?{\kern\digitwidth}
\halign{
\hfil#\hfil & \quad#\hfil && \quad\hfil#\hfil \cr
JD & \hfil Date & Instr. & Int. time & Wavel. range & FWHM \cr
\noalign{\vskip -4pt}
$-$2440000 & \hfil (1990) & config. & (s) & (\AA) & (arcsec) \cr
\noalign{\medskip}
7894.75 & ?2 Jan. & A & 1700 & 6040--6850 & 1.8 \cr
7897.73 & ?5 Jan. & A & 1300 & 6030--6840 & 1.4 \cr
7902.72 & 10 Jan. & A & 1300 & 6030--6840 & 1.5 \cr
7911.73 & 19 Jan. & B & 1000 & 6100--6900 & 2.6 \cr
7911.77 & 19 Jan. & C & ?600 & 6040--6870 & 2.8 \cr
7916.77 & 24 Jan. & C & ?350 & 6040--6870 & 2.4 \cr
7920.74 & 28 Jan. & B & 1000 & 6100--6900 & 1.7 \cr
7923.73 & 31 Jan. & C & ?400 & 6100--6930 & 1.3 \cr
7928.77 & ?5 Feb. & B & 1000 & 6100--6900 & 1.8 \cr
7931.73 & ?8 Feb. & B & 1000 & 6110--6910 & 1.2 \cr
7934.60 & 11 Feb. & B & 1000 & 6100--6900 & 3.0 \cr
7939.71 & 16 Feb. & C & ?400 & 6000--6820 & 1.2 \cr
7944.73 & 21 Feb. & A & 1000 & 6060--6870 & 3.6 \cr
7957.68 & ?6 Mar. & A & 1000+1000 & 6050--6870 & 1.9, 2.1 \cr
7968.69 & 17 Mar. & C & 1000 & 5960--6790 & 2.1 \cr
7975.68 & 24 Mar. & C & ?600 & 6050--6880 & 1.8 \cr
7984.53 & ?2 Apr. & C & ?350 & 6050--6880 & 1.0 \cr
7995.65 & 13 Apr. & B & 1000 & 6080--6880 & 2.6 \cr
7996.61 & 14 Apr. & B & 1000 & 6080--6880 & 1.6 \cr
7997.62 & 15 Apr. & B & 1500 & 6090--6890 & 3.6 \cr
8014.50 & ?2 May  & C & ?350 & 6060--6890 & 1.3 \cr
8025.52 & 13 May  & C & ?350 & 6070--6890 & 1.4 \cr
8027.66 & 15 May  & A & 1000 & 6040--6850 & 1.3 \cr
8031.63 & 19 May  & A & 1000 & 6040--6850 & 0.8 \cr
8036.58 & 24 May  & C & ?350 & 6070--6890 & 1.6 \cr
8045.52 & ?2 June & B & 1000 & 6100--6900 & 1.3 \cr
}}$$
\vfill\eject
\null\vfill
\centerline{{\bf Table 2.} Journal of observations for \mkn: \hb\ spectra}
\medskip
$$\vbox{
\newdimen\digitwidth
\setbox0=\hbox{\rm0}
\digitwidth=\wd0
\catcode`?=\active
\def?{\kern\digitwidth}
\halign{
\hfil#\hfil & \quad#\hfil && \quad\hfil#\hfil \cr
JD & \hfil Date & Instr. & Int. time & Wavel. range & FWHM \cr
\noalign{\vskip -4pt}
$-$2440000 & \hfil (1990) & config. & (s) & (\AA) & (arcsec) \cr
\noalign{\medskip}
7923.74 & 31 Jan. & C & ?700 & 4460--5280 & 1.2 \cr
7934.62 & 11 Feb. & B & 2000 & 4450--5250 & 2.1 \cr
7984.53 & ?2 Apr. & D & ?600 & 4460--5250 & 1.2 \cr
7996.63 & 14 Apr. & B & 1200 & 4440--5230 & 1.6 \cr
7997.65 & 15 Apr. & B & 1500 & 4440--5230 & 3.5 \cr
8014.52 & ?2 May  & C & ?700 & 4420--5240 & 1.0 \cr
8027.68 & 15 May  & A & 2000 & 4390--5200 & 1.5 \cr
8036.59 & 24 May  & C & ?700 & 4430--5240 & 1.5 \cr
8047.52 & ?4 June & B & 2000+2000 & 4440--5240 & 1.1, 1.3 \cr
}}$$
\vfill\eject
\null\vfill
\centerline{{\bf Table 3.} Instrumental configurations used for the LAG
spectroscopic campaign}
\medskip
$$\vbox{
\halign{
\hfil#\hfil & \ts\hfil#\hfil & \ts#\hfil &&
\ts\hfil#\hfil \cr
Symb. & Tel. & Instr. & Cam. & CCD & Pixel & Scale$_\perp$ & Disp. &
Resol. \cr
\noalign{\vskip -4pt}
&&&&& size \cr
\noalign{\vskip -4pt}
&&& (mm) && ($\mu$m) & (arcsec pxl$^{-1}$) & (\AA\ mm$^{-1}$) & (\AA) \cr
\noalign{\medskip}
A & INT & IDS           & 235 & GEC & 22.0 & 0.65 & 66.5 & 2.9 \cr
B & INT & IDS           & 500 & GEC & 22.0 & 0.30 & 66.1 & 7.3 \cr
C & WHT & ISIS$_{red}$  & 500 & EEV & 22.5 & 0.34 & 33.0 & 1.5 \cr
D & WHT & ISIS$_{blue}$ & 500 & GEC & 22.0 & 0.33 & 64.0 & 2.8 \cr
}}$$
\vfill\eject

\null\vfill
\centerline{{\bf Table 4.} Continuum and broad \ha\ light curves}
\medskip
$$\vbox{
\halign{
\hfil#\hfil && \quad\hfil#\hfil \cr
JD & $^a$F$_{\lambda6563}$ & $^a$F$_{\lambda6563}^{corr}$ &
$^a\sigma_{F_{\lambda6563}}$ & $^b$F(b\ha) & $^b\sigma_{F(bH\alpha)}$ \cr
\noalign{\vskip -4pt}
$-$2440000 \cr
\noalign{\medskip}
7894.75 & 6.45 & 6.22 & 0.19 & 2.26 & 0.07 \cr
7897.73 & 6.36 & 6.28 & 0.19 & 2.36 & 0.07 \cr
7902.72 & 6.52 & 6.41 & 0.19 & 2.40 & 0.07 \cr
7911.73 & 6.36 & 5.83 & 0.19 & 2.27 & 0.07 \cr
7911.77 & 6.45 & 5.85 & 0.19 & 2.31 & 0.07 \cr
7916.77 & 6.32 & 5.87 & 0.19 & 2.35 & 0.07 \cr
7920.74 & 6.43 & 6.24 & 0.19 & 2.61 & 0.08 \cr
7923.73 & 5.96 & 5.93 & 0.18 & 2.53 & 0.07 \cr
7928.77 & 6.14 & 5.91 & 0.18 & 2.51 & 0.07 \cr
7931.73 & 6.27 & 6.28 & 0.19 & 2.67 & 0.08 \cr
7934.60 & 6.52 & 5.86 & 0.26 & 2.47 & 0.10 \cr
7939.71 & 5.60 & 5.61 & 0.17 & 2.53 & 0.07 \cr
7944.73 & 5.99 & 5.13 & 0.30 & 2.32 & 0.11 \cr
7957.68 & 5.45 & 5.19 & 0.44 & 2.22 & 0.18 \cr
7968.69 & 5.36 & 5.02 & 0.27 & 2.26 & 0.11 \cr
7975.68 & 4.71 & 4.48 & 0.56 & 2.13 & 0.26 \cr
7984.53 & 5.06 & 5.15 & 0.15 & 2.32 & 0.07 \cr
7995.65 & 5.45 & 4.92 & 0.16 & 2.35 & 0.07 \cr
7996.61 & 5.22 & 5.06 & 0.16 & 2.40 & 0.07 \cr
7997.62 & 5.73 & 4.87 & 0.23 & 2.35 & 0.09 \cr
8014.50 & 4.55 & 4.51 & 0.14 & 2.21 & 0.07 \cr
8025.52 & 4.58 & 4.51 & 0.14 & 2.08 & 0.06 \cr
8027.66 & 4.40 & 4.37 & 0.13 & 2.14 & 0.06 \cr
8031.63 & 4.15 & 4.32 & 0.12 & 2.15 & 0.06 \cr
8036.58 & 4.52 & 4.37 & 0.13 & 2.11 & 0.06 \cr
8045.52 & 4.60 & 4.56 & 0.14 & 2.15 & 0.06 \cr
\noalign{\medskip}
\multispan6{$^a$\ 10$^{-15}$ \ergscma\hfil}\cr
\multispan6{$^b$\ 10$^{-12}$ \ergscm\hfil}\cr
}}$$
\vfill\eject

\null\vfill
\centerline{{\bf Table 5.} Broad \hb\ light curve}
\medskip
$$\vbox{
\halign{
\hfil#\hfil && \quad\hfil#\hfil \cr
JD & $^a$F(b\hb) & $^a\sigma_{F(bH\beta)}$ \cr
\noalign{\vskip -4pt}
$-$2440000 \cr
\noalign{\medskip}
7923.74 & 0.737 & 0.037 \cr
7934.62 & 0.738 & 0.037 \cr
7984.52 & 0.701 & 0.035 \cr
7996.63 & 0.593 & 0.030 \cr
7997.65 & 0.589 & 0.029 \cr
8014.52 & 0.534 & 0.027 \cr
8027.68 & 0.485 & 0.024 \cr
8036.59 & 0.507 & 0.025 \cr
8047.52 & 0.470 & 0.023 \cr
\noalign{\medskip}
\multispan3{$^a$\ 10$^{-12}$ \ergscm\hfil}\cr
}}$$
\bye